\documentclass[prl,twocolumn,superscriptaddress,floatfix]{revtex4}

\usepackage{epsfig,dcolumn,amsmath,latexsym}

\usepackage{subfigure}
\usepackage[normalem]{ulem}
\usepackage{cancel}
\usepackage{multirow}
\usepackage{color}

\definecolor{red}{rgb}{1,0,0}
\definecolor{blue}{rgb}{0,0,1}
\definecolor{black}{rgb}{0,0,0}

\def\pd343{Ce$_3$Pd$_3$Bi$_4$}
\def\pt343{Ce$_3$Pt$_3$Bi$_4$}
\def\px343{Ce$_3$Pd$_{3-x}$Pt$_x$Bi$_4$}

\begin{document}

\title{From Trivial Kondo Insulator \pt343 to Topological Nodal-line Semimetal \pd343} \preprint{1}

\author{Chao Cao}
 \email[E-mail address: ]{ccao@hznu.edu.cn}
 \affiliation{Condensed Matter Group,
  Department of Physics, Hangzhou Normal University, Hangzhou 310036, P. R. China}
 \affiliation{Center of Correlated Materials, Zhejiang University, Hangzhou 310058, China}

\author{Guo-Xiang Zhi}
 \affiliation{Department of Physics, Zhejiang University, Hangzhou 310013, P. R. China}
 
\author{Jian-Xin Zhu}
 \email[E-mail address: ]{jxzhu@lanl.gov}
 \affiliation{Theoretical Division and Center for Integrated Nanotechnologies, Los Alamos National Laboratory, Los Alamos, New Mexico 87545, USA}
\date{\today}

\begin{abstract}
Using the density functional theory combined with dynamical mean-field theory, we have performed systematic study of the electronic structure and its band topology properties of \pt343 and \pd343. At high temperatures ($\sim$290K), the electronic structures of both compounds resemble the open-core 4$f$ density functional calculation results. For \pt343, clear hybridization gap can be observed below 72K, and its coherent momentum-resolved spectral function below 18K exhibits an topologically trivial indirect gap of $\sim$6 meV and resembles density functional band structure with itinerant 4$f$ state. For \pd343, no clear hybridization gap can be observed down to 4K, and its momentum-resolved spectral function resembles electron-doped open-core 4$f$ density functional calculations. The band nodal points of \pd343 at 4K are protected by the gliding-mirror symmetry and form ring-like structure. Therefore, the \pt343 compound is topologically trivial Kondo insulator while the \pd343 compound is topological nodal-line semimetal.
\end{abstract}

\maketitle


The electronic structure of strongly correlated systems is one of the most intriguing problems in condensed matter physics. Taking heavy fermion systems as an example, its electronic structure exhibits strong temperature dependence and Kondo effect, due to the subtle interplay  between the hybridization of $f$-electrons with the itinerant band electrons and strong Coulomb repulsion among the $f$-electrons.  Recently, there are renewed interests in these systems due to the discovery of electron band topology~\cite{TI:KaneRMP,TI:ZhangRMP}. In addition to possible topological Kondo insulators (TKIs)~\cite{TI:SmB6NM,MKI:Coleman}, the possibility of Weyl Kondo semimetals WKSs has been brought up~\cite{WKS:QMSi,CeSb:Guo,PhysRevB.97.155134,YbPtBi:Guo,PhysRevLett.118.246601}.  In particular, the proposal of a truly heavy-fermion WKS~\cite{WKS:QMSi}  has stimulated intensive theoretical and experimental studies~\cite{PhysRevB.97.155134,YbPtBi:Guo,PhysRevLett.118.246601}. Experimentally, the realization of a TKI-WKS phase transition from~\pt343 to \pd343 has recently been hotly discussed~\cite{PhysRevLett.118.246601,2018arXiv181102819D,Kushwara:2019}. In this Letter, using state-of-art first-principles method based on density functional theory (DFT) and its combination with dynamical mean-field theory (DMFT)~\cite{method:dmft1,method:dmft2,RevModPhys.83.349}, we show that the \pt343 is a topologically trivial Kondo insulator; while \pd343 is a topological nodal-line Kondo semimetal, which is protected by the non-symmorphic symmetry of the crystal.

The DFT calculations were performed using plane-wave projected augmented wave method as implemented in the VASP~\cite{method:vasp,method:pawvasp} code, and cross-checked with both PWscf code~\cite{method:qe}  and the full-potential linearized augmented plane-wave (FP-LAPW) Wien2k code~\cite{method:wien2k}
Experimental lattice parameters were used for both compounds~\cite{pt343_structure,pd343_structure}. The DMFT calculations were performed using DMFTF package in connection with the Wien2k code~\cite{method:edmftf}, with $U$=6.0 eV and $J$=0.7 eV for Ce-4$f$ states. The continuous-time quantum Monte Carlo (CT-QMC) method was employed to solve the Anderson impurity problem~\cite{method:ctqmc_dmft}, and charge-density self-consistency was achieved. For more details about the DFT and DMFT calculations, please refer to the Supplementary Information (SI)~\cite{SupplementaryInfo}.

  \begin{figure*}[htp]
   \includegraphics[width=18cm]{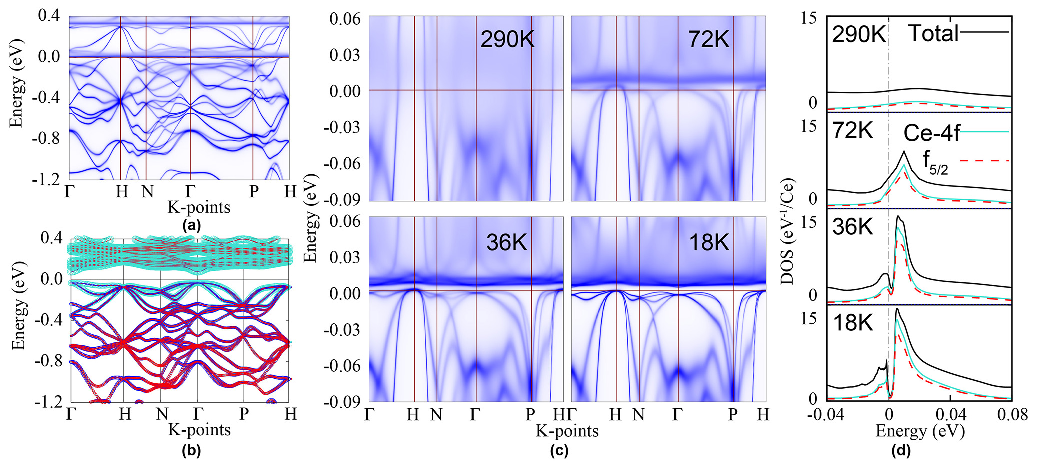}
    \caption{Electronic structure of \pt343. (a) The momentum-resolved spectral function from DFT+DMFT at 18 K. (b) The DFT band structure with itinerant Ce-4$f$ states. The contributions from Ce-4$f$, Bi-6$p$ and Pt-5$d$ states are represented by turquoise, blue and red circles, respectively. (c) The DFT+DMFT momentum-resolved spectral functions at 290 K, 72 K, 36 K and 18 K. (d) The DFT+DMFT spectral density at 290 K, 72 K, 36 K and 18 K. The solid black line is total spectral density, while the turquoise solid line, red dashed line, and blue dotted line represent Ce-4$f$, Ce-4$f_{5/2}$, and  Ce-4$f_{7/2}$ contributions, respectively.\label{fig:pt343}}
  \end{figure*}

Both \pt343 and \pd343 compounds are body-centered cubic crystals with symmetry group $I\bar{4}3d$ (No. 220), which contains 6 gliding mirror symmetry operations. Our main results of \pt343 is shown in Fig. \ref{fig:pt343}. At high temperatures ($T$=290 K), the Ce-4$f$ electrons are localized, leading to incoherent $f$-bands near the Fermi energy $E_F$, while coherent conduction bands can still be identified between $\Gamma$-H, H-N, and P-H (Fig.~\ref{fig:pt343}c), thus the system is metallic. The spectral density shows two small humps near the Fermi level, formed by the Ce-4$f_{5/2}$ at $E_F$ and Ce-4$f_{7/2}$ 380 meV above the Fermi energy, respectively.  As the temperature is lowered to 72 K, the flat Ce-4$f_{5/2}$ band begin to form around 10 meV above $E_F$, which strongly hybridizes with the conduction bands. The spectral density shows sharper Ce-4$f$ peak, but without any trace of hybridization gap at 72 K. At 36 K, the hybridization gap is already present, as the spectral density shows a clear dip around the Fermi level $E_F$. At and below 36K, the many-body self-energy of Ce-4$f_{5/2}$ states within the energy gap shows a Fermi-liquid-like behavior $\mathrm{Im}[\Sigma_{5/2}(\omega)]\approx\alpha(\omega-\omega_0)^2+\Sigma_0$ with a very small $\Sigma_0$ and $\omega_0$ inside the gap (Refer to  Figs.~S-2 and S-3 in SI for details), while such fitting is not possible at 72 K. Therefore, the coherence temperature is expected to be between 36K and 72K from our calculation. Experimentally, the hybridization gap starts to form below 100 K, and its maximum value is measured to be approximately 50 K (or 4.3 meV)~\cite{PhysRevLett.72.522,PhysRevB.42.6842,PhysRevB.44.6832,PhysRevB.55.7533,PhysRevB.94.035127}. In addition, the momentum-resolved spectral function clearly shows an energy gap at $E_F$, and an indirect gap of $\sim$ 6 meV can be identified from integrated and momentum-resolved spectral function plots at 18 K. In all these DFT+DMFT calculations, the Ce-4$f_{5/2}$ occupation ranges from around 1.00 at 290 K to 0.98 at 18 K, suggesting negligible valence fluctuation. Therefore, the system is consistent with the Kondo lattice model, where the $f$-electrons can be treated as localized spins coupled to conduction electrons via a Kondo exchange interaction.

We notice that the coherent momentum-resolved spectral function at 18 K from DFT+DMFT calculation is indeed very similar to DFT results with itinerant Ce-4$f$ states (Fig. \ref{fig:pt343}b). The DFT band structure yields an insulating ground state with $\sim$140 meV indirect gap near $\Gamma$. The highest occupied state at $\Gamma$ is the four-fold degenerate $\Gamma_8$ state in DFT calculation. It splits into two doubly degenerate states along $\Gamma$-H line. At H point, these four states are joined by the doubly degenerate state from $\Gamma_7$ at around 0.2 eV below $E_F$ at $\Gamma$, as well as another doubly degenerate states stemming from the split $\Gamma_8$ state at $\sim$0.4 eV below $E_F$, to form an 8-fold degenerate state as the highest occupied state. This is exactly the same result we obtained for the coherent  momentum-resolved spectral function at 18 K from DFT+DMFT calculations. Such analysis can be conducted along all high symmetry lines we presented, although the band energies are strongly renormalized in the DFT+DMFT calculations.  The DFT calculation yields an insulating ground state with an indirect gap of $\sim$ 140 meV, which is about 20 times larger of the DFT+DMFT result at 18 K. This is consistent with the mass enhancement factor $m^*/m\approx26$ obtained from the many-body self-energy $\Sigma$.

The striking similarities between the DFT result and the DFT+DMFT result below the coherence temperature suggest that the coherent DFT+DMFT momentum-resolved spectral function can be adiabatically transformed from DFT result without closing the energy gap. Therefore, the topological properties of \pt343 coherent spectral function from the DFT+DMFT is anticipated to be the same as its DFT band structure. It is worth noting that the global energy gap is already present in the calculations for \pt343 even when the spin-orbit coupling is turned off (SI Fig. S-4a), and is therefore not related to the band inversion. Therefore, the \pt343 is expected to be a Kondo insulator with trivial band topology. In fact, we have fitted the DFT band structure to a tight-binding Hamiltonian using the maximally localized Wannier function method~\cite{method:mlwf}(Refer to SI Fig. S-6(a)), and performed Wilson loop calculations~\cite{method:wilsonloop}. The resulting $\mathcal{Z}_2$ is (0;000) for \pt343 (Refer to  SI Fig. S-8(a)), showing its trivial topology nature. In addition, we have also employed the so-called topological Hamiltonian, i.e. inverse of zero-frequency Green's function~\cite{PhysRevX.2.031008,PhysRevB.85.165126,Wang_2013} to determine the same property for the correlated many-body Hamiltonian from DFT+DMFT. The resulting $\mathcal{Z}_2$ is also (0;000) as expected from the adiabatic argument (Refer to SI Fig. S-8(b)). Therefore, the experimentally observed resistivity saturation must have an origin other than topological arguments, and robust surface states are not guaranteed. Indeed, the Hall measurement suggested the conduction of \pt343 at low temperature is dominated by bulk in-gap states, unlike SmB$_6$, and is susceptible to low concentrations of disorders~\cite{PhysRevB.94.035127}. 

  \begin{figure}[htp]
   \includegraphics[width=8cm]{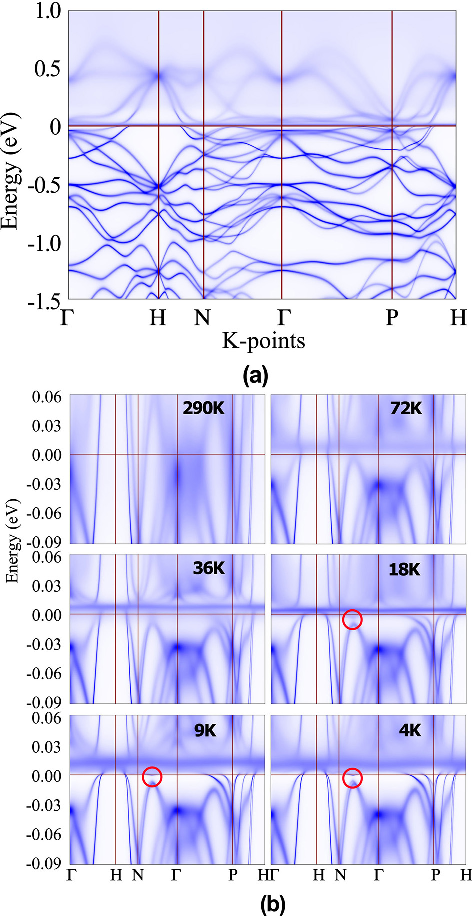}
    \caption{Electronic structure of \pd343. (a) The DFT+DMFT momentum-resolved spectral function in the energy range [-1.5,1.0] eV with respect to the Fermi energy at 4 K. (b) The DFT+DMFT momentum-resolved spectral function in a zoomed-in energy range close to the Fermi level at 290 K, 72 K, 36 K, 18 K, 9 K and 4 K. \label{fig:dmft_pd343}}
  \end{figure}

Having established that the \pt343 is a trivial Kondo insulator, we now focus on \pd343 compound. The \pd343 compound is isostructural to \pt343 with nearly identical lattice constants and slightly different Bi atomic coordinates. It is therefore tempting to believe that the electronic structure of \pd343 is similar to \pt343. However, our DFT+DMFT calculations show that the temperature dependence of \pd343 electronic structure is very different (Fig.~\ref{fig:dmft_pd343}). At 290 K, the Ce-4$f$ states are localized, as suggested by the highly dispersive quasi-particle conduction bands in Fig.~\ref{fig:dmft_pd343}b. As the temperature is decreased, the Ce-4$f$ flat band starts to form slightly above the Fermi level. However, they never form coherent Ce-4$f$ states down to 4K, indicated by its large imaginary part of self-energy close to the Fermi level ($\approx$ 30 meV at 4K. See also SI Figs. S-2 and S-3). The ligand state near the Fermi level, on the other hand, becomes sharp below 9K. This is most prominent from the states between $\Gamma$-N (indicated by the red circles in Fig.~\ref{fig:dmft_pd343}, also in SI Fig. S-9) and $\Gamma$-P near $E_F$. During the whole temperature range, we do not observe well defined gap opening at the Fermi level down to 4 K in our calculation, thus \pd343 is most likely to be a correlated metal at low temperatures and the Ce-4$f$ electrons remain localized at low temperatures for \pd343.

\begin{figure}[htp]
   \includegraphics[width=8cm]{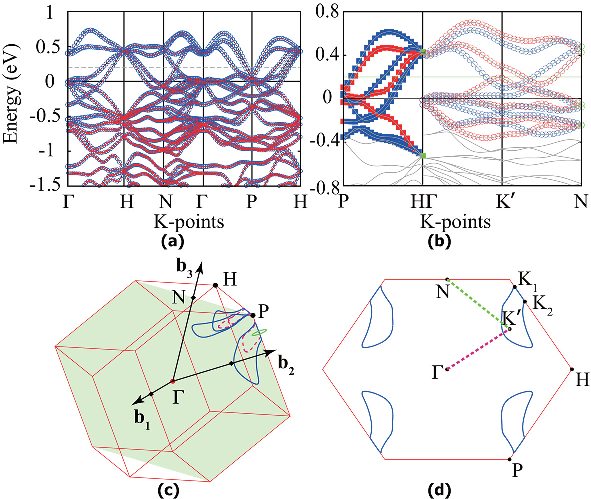}
    \caption{DFT electronic structure of \pd343. (a) The DFT band structure calculated by assuming open-core Ce-4$f$ state. The size of the red circles are proportional to the Pd-4$d$ state contributions, while the blue circles are contributions from Ce-4$f$ and Bi-6$p$ states.  (b) The band structure on the gliding mirror plane $\Gamma$-N-P. K$^{\prime}$ is (0, 1/4, 1/4). Along P-H, the bands can be classified with C$_3$ eigenvalues $e^{\pm i2\pi/3}$ (red squares) and 1 (blue squares). In other directions, the bands can be labeled with $-i$ (red circles) or $i$ (blue circles). (c) The first Brillouin zone showing the high symmetry points and the nodal rings. The green circle is the path used to calculate the Berry phase. (d) The nodal rings (blue lines) on one of the gliding-mirror planes. \label{fig:dft_pd343}}
  \end{figure}

We now compare the DFT+DMFT momentum-resolved spectral function for \pd343 at 4 K with DFT calculations treating Ce-4$f$ state as core-electrons. Similar results can be obtained using LDA+$U$ calculations with proper treatment of time-reversal symmetry (See SI FIG. S-5(c) and (d) for details). Again, the momentum-resolved spectral function at 4 K for \pd343 resembles the DFT results, if the Fermi level in the DFT band structure is shifted $\sim$ 0.2 eV higher (marked by the gray line in Fig.~\ref{fig:dft_pd343}(a)-(b)). The DFT band structure has a large band gap of nearly 500 meV at approximately 1 eV above the Fermi level, or equivalent to counting 6 more electrons per unit cell. The states near $E_F$ are dominated by Bi-6$p$ and Ce-5$d$ orbitals with a small mixture of the Pd-4$d$ orbitals. We note that the centers of gravity of the Pd-4$d$ orbitals are located in the energy region between [-4, -2] eV below $E_F$. The highest occupied states at $\Gamma$ are doubly degenerate $\Gamma_7$, quarterly degenerate $\Gamma_8$ and doubly degenerate $\Gamma_6$ (from higher energy to lower energy), respectively. Along $\Gamma$-H line, the $\Gamma_8$ states splits into two doubly degenerate bands, and both $\Gamma_6$ and $\Gamma_7$ states remain doubly degenerate. The $\Gamma_7$ state and one pair of the split $\Gamma_8$ state are joined by other 2 doubly degenerate bands to form an 8-fold degenerate highest occupied state at H. It again splits into 4 doubly degenerate states along H-N, and further splits into 8 singly degenerate states along N-$\Gamma$. Such splitting are all present in 4 K momentum-resolved spectral function from DFT+DMFT calculations.

Similar to our previous analysis for \pt343, we argue that the similarity between 4 K momentum-resolved spectral function and DFT band structure ensures that the topological properties of \pd343 can be analyzed using the DFT results with proper Fermi level shifting. Judging from the bands crossing the Fermi level between N-$\Gamma$ and around P, the DFT results can be compared with DFT+DMFT results (see also SI Fig. S-9(a)) if the DFT Fermi level is shifted upward by approximately 0.2 eV. In addition, the Fermi surface from open-core Ce-4$f$ DFT calculations matches that from the DFT+DMFT calculations at 4 K after 0.2 eV Fermi level shifting, both of which consist of 4 sheets around H and 4 sheets around P points (SI Fig. S-10(a) and (c)). On the contrary, the Fermi surface from open-core Ce-4$f$ DFT calculations without Fermi level shifting consists of 2 sheets around H and 2 sheets around P points (SI Fig. S-10(b)). From the DFT DOS (SI Fig. S-7(b)), we found that 0.2 eV Fermi level shifting is equivalent to 1 electron doping per formula (or 2e per unit cell). Therefore, we need to locate for the nodal structures between band 4 and band 5 (band indices are marked on SI Fig. S-11(b)). Notice that the crystal contains 6 gliding-mirror planes in \{100\} and five other equivalent directions, the states in $k_x=k_y$ and all equivalent k-planes can be classified using the mirror symmetry. Let us consider 1 of the 6 equivalent planes ($\eta$, $\xi$, $\xi$) (in crystal coordinate). First of all, there is a high symmetry line P-H within this plane, which is shared by other 2 gliding mirror planes. In addition to the gliding mirror symmetry, $C_3$ rotation symmetry is also preserved along this line. Therefore, all states can be labelled with $C_3$ eigenvalues $e^{\pm i2\pi/3}$ or 1, the former is doubly degenerate. At point P, bands 2-4 and 5-7 are triply degenerate due to the local T$_d$ symmetry. As it moves along P-H, they both split into a singly degenerate band (4 and 7) and a pair of doubly degenerate band (2,3 and 5,6). Band 4 then crosses doubly degenerate band 5/6 twice, creating two triply degenerate points at K$_1$ (0.238, 0.254, 0.254) (all the coordinates are in direct unit, unless otherwise specified) and K$_2$ (0.046, 0.318, 0.318), as well as a segment of nodal-line between them. It is also instructive to notice that bands 1-7 and 10 will join together at H and form an 8-fold degenerate state; while bands 8-9 are doubly degenerate states and will be joined by other 6 states at H to form another 8-fold degenerate state.  Secondly, we take look at another K-path from $\Gamma$ to K$^{\prime}$ (0.0, 0.25, 0.25) point. As all these states are in the mirror plane, they can be labeled with $\pm i$. Evidently, there is a band inversion between the band 4 and 5 between $\Gamma$ and K$^{\prime}$ points, and therefore there must be a nodal point. In fact, such nodal point can be identified between K$^{\prime}$ and any K-point along $\Gamma$-P or $\Gamma$-H. Therefore, we conclude that these nodal points must form a nodal ring within the triangle $\Gamma$-H-P. Since there are 6 gliding-mirror planes, and each plane contains 4 such triangles within the first BZ, there are 24 nodal rings, forming 8 groups around 8 P points, each group contains 3 nodal rings sharing the same K$_1$-K$_2$ nodal line. The Berry phase around a circular path perpendicular to P-H around the K$_1$-K$_2$ nodal line is $\pi$ or $-\pi$, indicating its nontrivial topology. In the open-core Ce-4$f$ electron DFT calculations, these nodal rings occupy energy ranges from 0.1 eV to 0.3 eV above $E_F$, therefore they cross the Fermi level if the Fermi level is shifted 0.2 eV upward (1 e doping per formula). In the DFT+DMFT calculations, these nodal rings can also be identified in the momentum-resolved spectral function at 4 K. From P-H, the doubly degenerate states 5/6 are not very sharp a few meVs below $E_F$. However, such crossings can still be identified at $\sim$ 50 meV below $E_F$. We have also employed the topological Hamiltonian method to determine the nodal structure of \pd343 at 4K. These results are shown in Fig. \ref{fig:dft_pd343}c with dashed magenta lines. The topological Hamiltonian yields nodal rings similar to the open-core DFT calculations, but the size of these rings are much reduced due to the many-body renormalization effect. In addition to the 24 nodal rings we described above, we have also identified band crossings at 6 H points, as well as at 24 points at (0.328, 0.153, 0.000) (in cartesian coordinates, \AA$^{-1}$ unit) and its symmetrically equivalent positions. However, the chirality of all these crossings are 0, meaning that they are not topological. 

An important question is what leads to the difference between \pt343 and \pd343 compounds. Since Pt and Pd belong to the same element family, and both compounds are isostructural with nearly identical lattice constants, the spin-orbit coupling (SOC) change seems to be one very plausible reason. However, if we compare the band states in the energy range [-0.5,1.0] eV in the DFT calculations by assuming open-core Ce-4$f$ electrons (SI Fig. S-5), it is clear that the \pt343 SOC splitting near $E_F$ is not very different from \pd343. From the DFT calculations, the spin-orbit coupling constants $\lambda_{\mathrm{SO}}$ are 0.453/0.659/1.051 eV for Ce/Pt/Bi atoms in \pt343, and 0.454/0.324/1.074 eV for Ce/Pd/Bi atoms in \pd343, respectively. More importantly, we notice that the \pd343 compound would also be an insulator with a large band gap in the DFT calculations if Ce-4$f$ electrons are treated as being itinerant. Therefore, such difference is more likely related to a change in hybridization. In fact, the hybridization function $\Delta$ for 4$f_{5/2}$ exhibits completely different behavior for \pt343 and \pd343. If we compare the imaginary part of $\Delta_{f5/2}$ in Matsubara frequency domain for both compounds at 18 K (please refer to SI Fig. S-12a), it is clear that $\text{Im}[\Delta_{f5/2}]$ for \pt343 blows up below $\omega_n=40$ meV, consistent with a Kondo insulator behavior~\cite{Tomczak_2018}; while for \pd343, $\text{Im}[\Delta_{f5/2}]$ remains finite and reduces in the same frequency range. Correspondingly, the hybridization function in real-frequency axis $\Delta(\omega)$ exhibits singular peaks at the Fermi level for \pt343; while such peaks are absent for \pd343 (See also SI Fig. S-12(b)).

In conclusion, we have performed a systematic DFT and DFT+DMFT study of \pt343 and \pd343 compounds. The \pt343 compound is a trivial Kondo insulator with an indirect gap of $\sim$6 meV. The \pd343 compound is a correlated metal with topological nodal lines at low temperatures. The hybridization change is the main driving force for the above-mentioned transition.

\begin{acknowledgements}
The authors are grateful to Qimiao Si,  S. K. Kushwaha, P. F. S. Rosa, N. Harrison, J. Lawrence, Huiqiu Yuan, Jianhui Dai, and Frank Steglich for stimulating discussions. C.C also acknowledges the hospitality of Los Alamos National Laboratory, where this work was initiated. This work at Los Alamos was carried out under the auspices of the U.S. Department of Energy (DOE)  National Nuclear Security Administration under Contract No. 89233218CNA000001.  It was supported by NSFC 11874137 and 973 Project 2014CB648400  (C.C. \& G.-X.Z.),  and  U.S. DOE BES under LANL-E3B5 and, in part, by Center for Integrated Nanotechnologies, a DOE BES user facility, in partnership with LANL IC Program for computational resources (J.-X.Z.). The calculations were performed also on the High Performance Computing Cluster of Center of Correlated Matters at Zhejiang University, and Tianhe-2 Supercomputing Center.
\end{acknowledgements}


\end{document}


\title{Supplementary Information: From Trivial Kondo Insulator \pt343 to Nodal-line Semimetal \pd343} \preprint{1}

\author{Chao Cao}
 \email[E-mail address: ]{ccao@hznu.edu.cn}
 \affiliation{Condensed Matter Group,
  Department of Physics, Hangzhou Normal University, Hangzhou 310036, P. R. China}
 \affiliation{Center of Correlated Materials, Zhejiang University, Hangzhou 310058, China}
\author{Guo-Xiang Zhi}
 \affiliation{Department of Physics, Zhejiang University, Hangzhou 310013, P. R. China}
\author{Jian-Xin Zhu}
 \email[E-mail address: ]{jxzhu@lanl.gov}
 \affiliation{Theoretical Division and Center for Integrated Nanotechnologies, Los Alamos National Laboratory, Los Alamos, New Mexico 87545, USA}
\date{\today}

\maketitle

\section{Calculation details of DFT and  DFT+DMFT}

\begin{figure*}[h]
\includegraphics[width=16 cm]{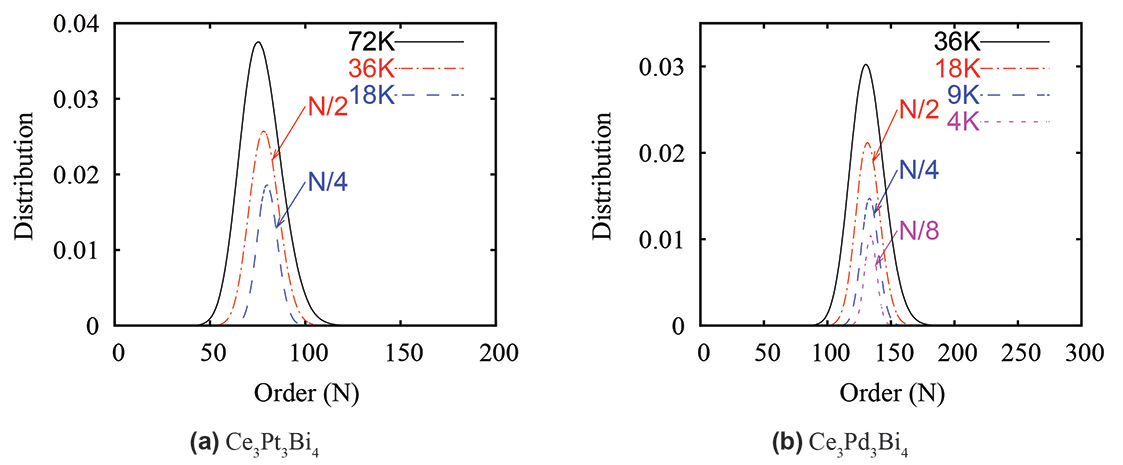}
  \caption{Distribution of perturbation orders for (a) \pt343 and (b) \pd343 at low temperature calculations. \label{fig:dmft_histogram}}
\end{figure*}

\begin{figure*}[htp]
 \includegraphics[width=16 cm]{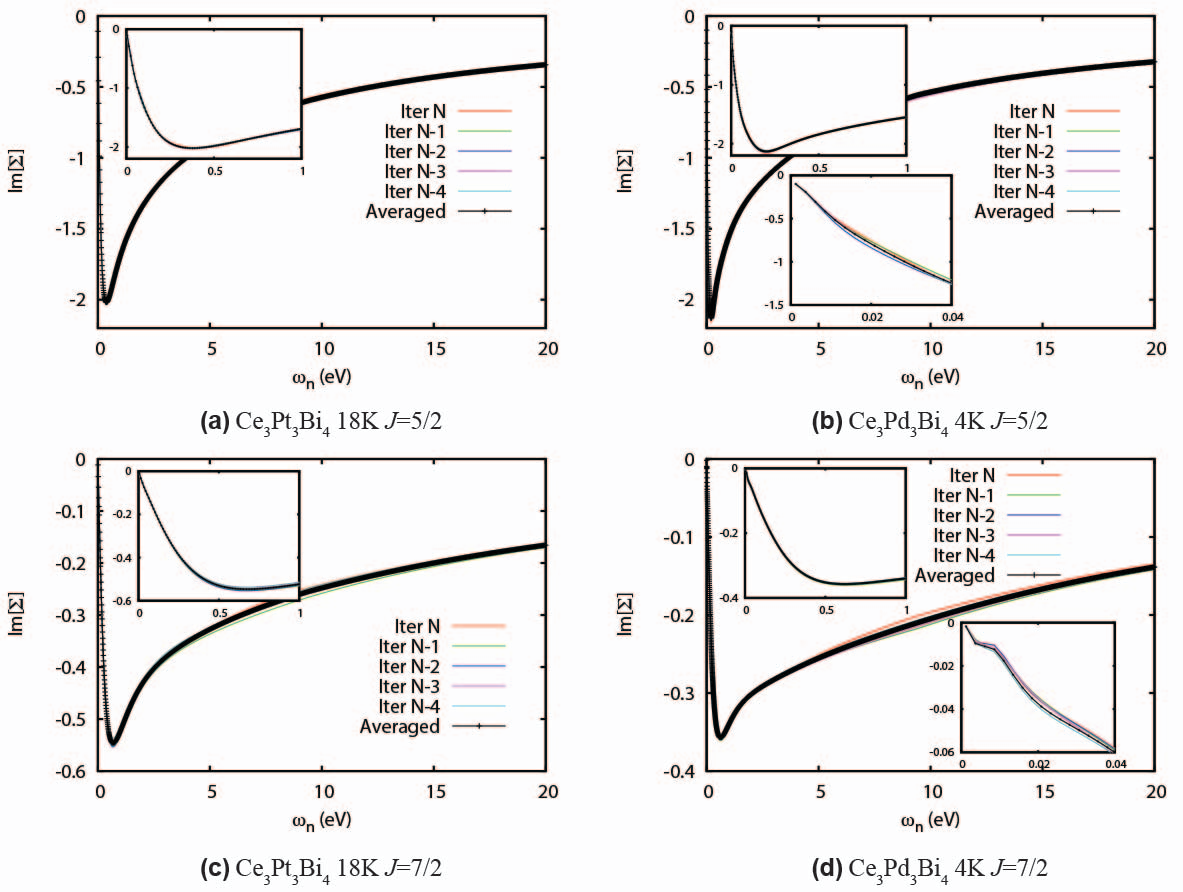}
  \caption{Imaginary part of self-energy $\Sigma(i\omega_n)$ in Matsubara frequency. (a) and (b) are for $J=5/2$ states, while (c) and (d) are for $J=7/2$ states. (a) and (c) are results for \pt343 at 18 K, while (b) and (d) are for \pd343 at 4 K. \label{fig:sigma_lowT}}
\end{figure*}

\begin{figure*}[htp]
 \includegraphics[width=16 cm]{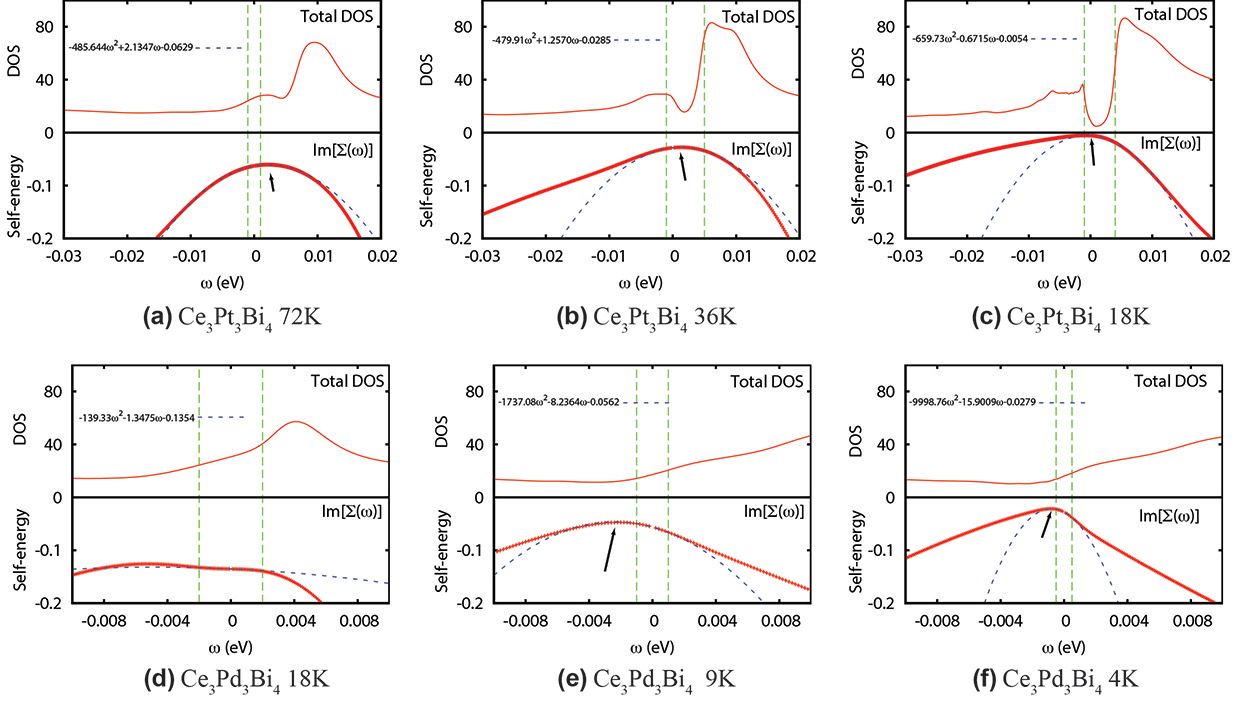}
  \caption{Integrated spectral function and imaginary part of self energy $\Sigma_{5/2}$ on real-frequency axis for (a-c) \pt343 at 72 K, 36 K, and 18 K and (d-f) \pd343 at 18 K, 9 K and 4 K. The dotted blue lines are the parabolic fittings to $\Sigma_{5/2}$ using least-square fitting method; the green lines denotes the fitting range; the black arrow indicates the position of maximum. The fitted function is also shown.\label{fig:sigma_real}}
\end{figure*}

The DFT calculations were performed using plane-wave projected augmented wave method as implemented in the VASP~\cite{method:vasp,method:pawvasp} code, and cross-checked with both PWscf code~\cite{method:qe}  and the full-potential linearized augmented plane-wave (FP-LAPW) Wien2k code~\cite{method:wien2k}. The plane-wave basis energy cut-off were chosen to be 480 eV in VASP calculations and 48 Ry in PWscf calculations. In FP-LAPW calculations, $RK_{\mathrm{max}}$ was set to 9. In all these calculations, a $12\times12\times12$ dense $\Gamma$-centered K-mesh was used to perform Brillouin zone integration. The VASP band structures were used to fit a tight-binding Hamiltonian using 108 (192) Wannier orbitals~\cite{method:mlwf} for DFT calculations with open-core (itinerant) Ce-4$f$ states, which were used to perform topology analysis by the Wilson loop method. 

For the DFT+DMFT calculations, we have employed $U$=6.0 eV and $J$=0.7 eV (corresponding to $F^0$=6.0 eV, $F^2$=8.345 eV, $F^4$=5.5747 eV, $F^6$=4.1226 eV) for Ce-4$f$ orbitals and a $12\times12\times12$ $\Gamma$-centered K-mesh was used. The states within [-10.0, 10.0] eV with respect to the Fermi level are projected to obtain the low-energy Hamiltonian used in the DMFT calculations. There are $\sim$310 states within this range, including all Pd-4$d$/Pt-5$d$, Bi-6$p$, Ce-4$f$ and more than 100 other unoccupied states (including Ce-5$d$ orbitals). As the spin-orbit coupling of Ce-4$f$ orbitals are orders of magnitudes larger than the crystal-field splitting (CEF), we took the approximation that CEF is not considered at the impurity solver level. Nevertheless, the CEF is considered at the DMFT (lattice) level. In the continue time quantum Monte Carlo (CT-QMC) impurity solver, full Coulomb interaction matrix was employed, and the 4$f$ impurity occupations were restricted to within [0, 3]. A singular value decomposition basis expansion to the Green's function up to $l=25$ was used in the impurity solver to reduce the memory cost. Nominal double counting scheme with $n_f^0=1$ was chosen.

For 4 K \pd343 calculations, we used 144 CPU-cores and $2\times10^9$ QMC steps for each QMC run. The calculation converges after 60 DMFT iterations, and we used additional 5 DMFT iterations to average the resulting self-energy function. For other calculations, we used 96 CPU-cores and QMC steps ranging from $2\times10^8$ (for 290 K) to $12\times10^8$ (for 9 K) for each QMC run. They converge after 20-40 DMFT iterations, and we used additional 5 DMFT iterations to average the resulting self-energy function. We show the histogram of perturbation orders in Fig.~\ref{fig:dmft_histogram}. Please be noted that for some of these plots, the order ($N$) is rescaled to fit in the plotting range. For the \pd343 4 K calculations, we truncated the perturbation beyond order of 3200.

\section{Self energies from DFT+DMFT calculations}

We show in Fig.~\ref{fig:sigma_lowT} the imaginary part of the DMFT self-energy $\Sigma$ in Matsubara frequency for the lowest calculated temperatures, i.e. \pt343 at 18 K and \pd343 at 4 K. We show the resulting $\Sigma$ from the last 5 DMFT iterations and their average that was used in the numerical analytical continuation (maximum entropy) process. The insets zoom in the same plots close to the zero frequencies. For \pt343 at 18 K, the imaginary part of the self-energy extrapolates to 0 when $\omega_n\rightarrow0$; while for \pd343 at 4 K, it extrapolates to a finite value at $\omega_n\rightarrow0$. For both calculations, the self-energy from last 5 iterations can hardly be distinguished from their average, showing that the calculations are well converged. For $J=7/2$ states, they both cut through the origin as $\omega_n\rightarrow0$, but there is no $J=7/2$ state at $E_F$, so they are irrelevant to the low-energy excitations. For \pd343 at 4 K, the self-energy of $J=7/2$ state exhibits some numerical instability below $\omega_n$=0.02 eV. But again, $J=7/2$ states are unoccupied, so they do are irrelevant to the low-energy excitations, which we are interested in.

Using the maximum entropy method, we obtain the self-energy of \pt343 and \pd343 on the real-energy axis (Fig.~\ref{fig:sigma_real}), which are then used to obtain the spectral function. We show here the imaginary part of $\Sigma_{5/2}$ of \pt343 and \pd343 at their respective three lowest temperatures, as well as low-energy parabolic fittings using the least-square fitting method. For metallic systems, the fitting range is between [-$T$, $T$] ($T$ is the temperature); for gapped systems, the fitting is taken inside the gap. For \pt343, the low energy behavior of $\mathrm{Im}[\Sigma_{5/2}]$ can always be well fitted using a parabolic function $\alpha(\omega-\omega_0)^2+\Sigma_0$. However, at 72K its minimum is located outside the fitting range and the minimum value is much larger than the temperature. Therefore, its Ce-4$f$ electrons are incoherent at 72K. At 36K and 18K, the minimums are located well inside the gap, and the minimum value is comparable or smaller than the gap size. Since the Fermi level is arbitrarily determined inside the gap, it means that by realigning the Fermi level to its proper value, the self energy can always be written as $\mathrm{Im}[\Sigma_{5/2}(\omega)]\approx \alpha \omega^2+\Sigma_0$ with small $\Sigma_0$, i.e. the Fermi liquid form. Therefore, they are coherent at 36 K and 18 K. For \pd343, the low-energy behavior of $\Sigma_{5/2}$ deviates from parabolic function at 18 K. At 9 K and 4 K, such fitting can be done, but the fitted minimum is orders of magnitude larger than the temperature (-27.9 meV at 4 K). In addition, the position of the minimum is also located outside the fitting range (beyond the thermal error). For metals, the Fermi energy is determined by total number of electrons, and they cannot be arbitrarily determined. Therefore, they do not behave like the Fermi liquids, and are incoherent down to 4K. 

  \begin{figure*}[htp]
  \includegraphics[width=16 cm]{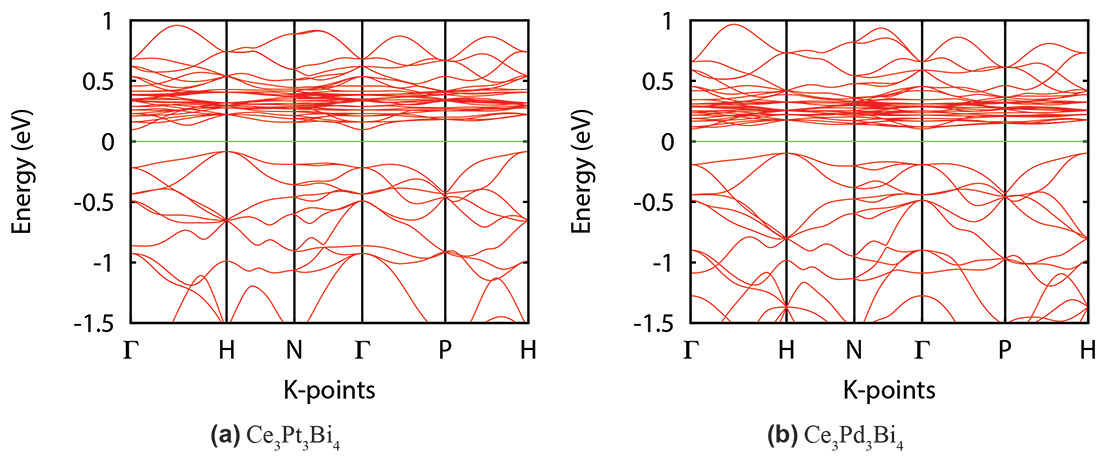}
    \caption{DFT band structure of (a) \pt343 and (b) \pd343 assuming itinerant $f$-electrons without spin-orbit coupling. The Fermi levels are aligned at 0. \label{fig:dft_itin}}
  \end{figure*}

Finally, the mass enhancement can be calculated using $m^{*}/m=Z^{-1}=1-\frac{\partial \mathrm{Re}[\Sigma(\omega)]}{\partial \omega} |_{\omega=0}$. It gives us a mass enhancement of approximately 26 and 65 for \pt343 at 18 K and \pd343 at 4 K, respectively.

\section{DFT band structures of $\mathbf{Ce}_3\mathbf{Pt}_3\mathbf{Bi}_4$ and $\mathbf{Ce}_3\mathbf{Pd}_3\mathbf{Bi}_4$ with itinerant and local $f$-electron states}

  \begin{figure*}[htp]
  \includegraphics[width=16 cm]{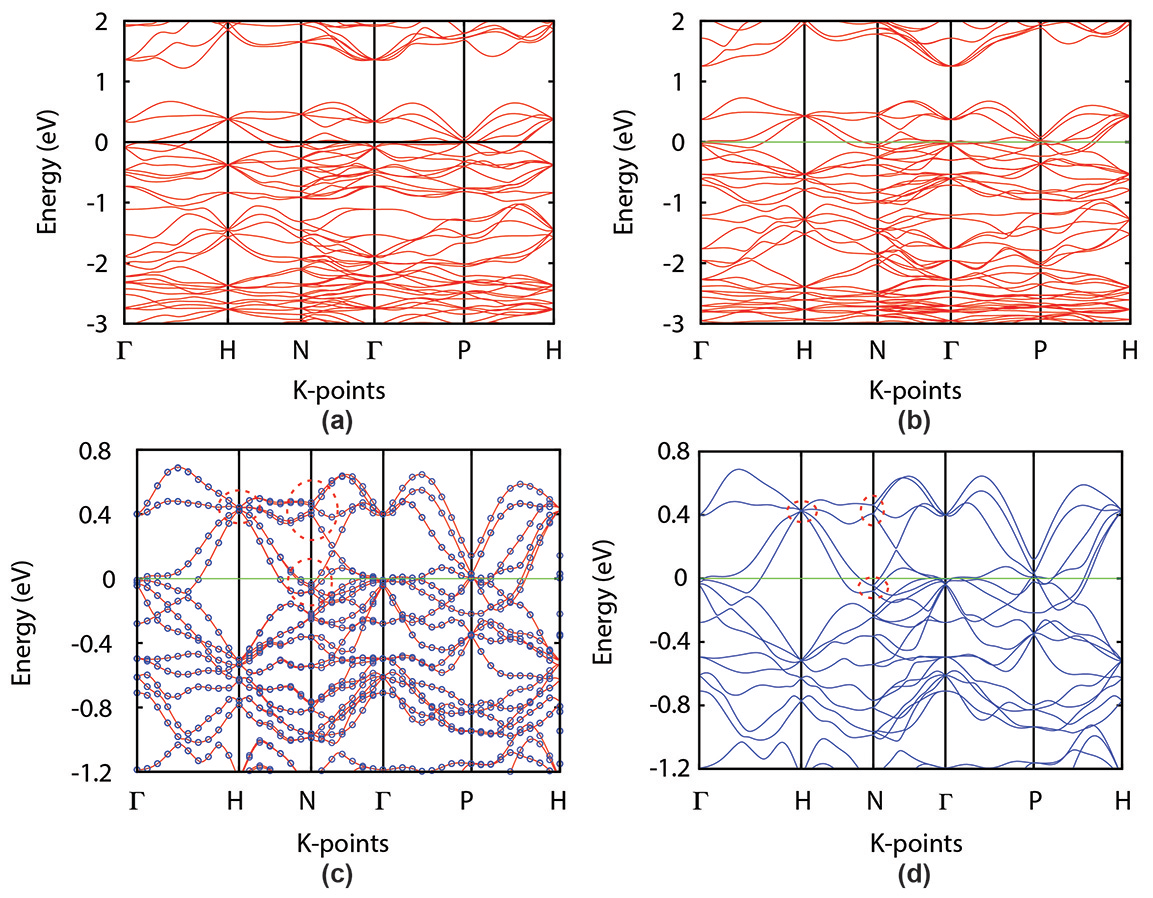}
    \caption{DFT band structure of (a) \pt343 and (b) \pd343 assuming local $f$-electrons with spin-orbit coupling. (c) \pd343 band structure using LDA+$U$ method. (d) Time-reversal symmetrized LDA+$U$ band structure for \pd343. In panel (c), the Wannier fitted tight-binding result is shown in blue circles, and the original DFT band structure is plotted using red solid lines. The Fermi levels are aligned at 0. \label{fig:dft_local}}
  \end{figure*}

  \begin{figure*}[htp]
 \includegraphics[width=18 cm]{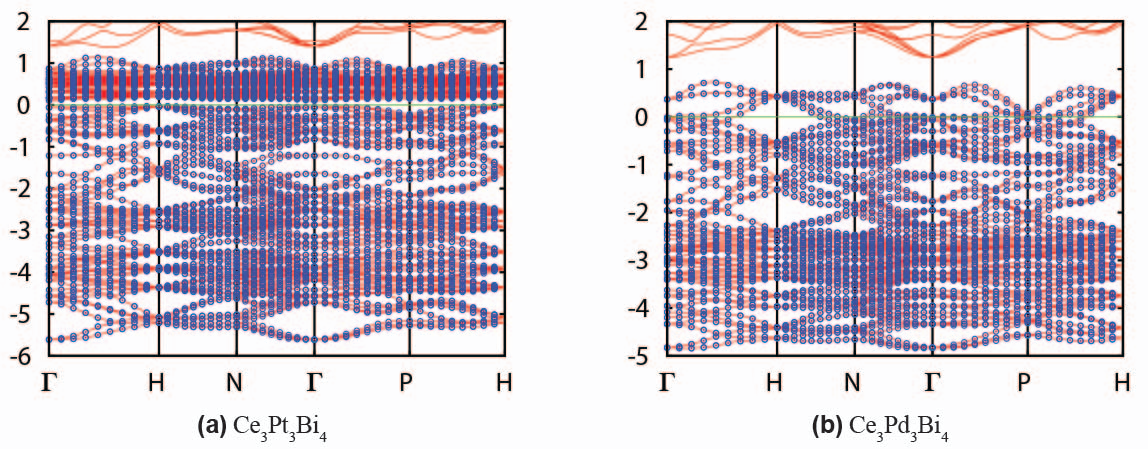}
    \caption{The comparison between Wannier orbital fitted band structure and DFT results with spin-orbit coupling. Both the band structure for (a) \pt343 assuming itinerant Ce-4f states and (b) \pd343 assuming local Ce-4f states are compared. The Fermi levels are aligned at 0. The fitted band structure using Wannier orbital TB model is plotted using blue circles, and the original DFT band structure is plotted using red lines. \label{fig:wann_fit}}
  \end{figure*}

In Fig.~\ref{fig:dft_itin}, we show the DFT band structure of \pt343 and \pd343 assuming itinerant $f$-electrons. Spin-orbit couplings (SOC) are not considered in these calculations. At itinerant $f$-electron DFT level, both systems are insulators with similar band gaps, no matter if SOCs is considered or not.  

\begin{figure*}[h]
  \includegraphics[width=18 cm]{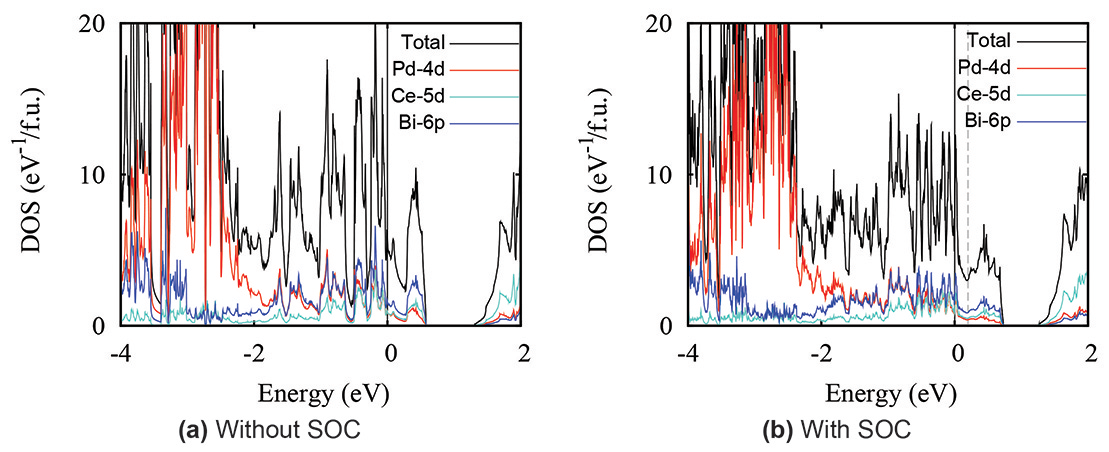}
  \caption{Density of states of \pd343 (a) without SOC and (b) with SOC. The Fermi level is aligned at 0. In the right panel, the grey line is 0.2 eV above $E_F$. \label{fig:dft_dos}}
\end{figure*}
 
\begin{figure*}[htp]
  \includegraphics[width=18 cm]{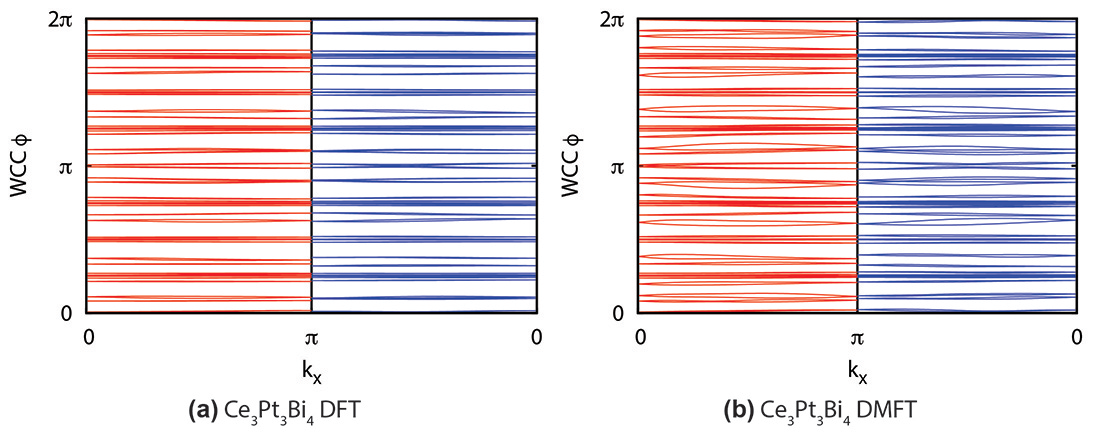}
  \caption{Wannier charge center flow chart of \pt343. (a) DFT result with itinerant $f$-electrons. (b) 18K DMFT result using zero-frequency Green's function. In either panel, the left half was $k_z$=0 plane and the right half was $k_z=\pi$ plane. \label{fig:wanncenter}}
\end{figure*}

In Fig.~\ref{fig:dft_local}(a) and (b), we show the DFT band structure of \pt343 and \pd343 assuming localized $f$-electrons. The SOCs are included as second-variational procedure. At localized $f$-electron DFT level, both systems are metallic. In Fig.~\ref{fig:dft_local}(c), we also show LDA+$U$ band structure of \pd343 compound, starting from nonmagnetic initial conditions. Overall speaking, the LDA+$U$ band structure show striking similarities to the open-core LDA calculations. However, the time-reversal symmetry is violated in our LDA+$U$ calculations, leading to spin-splittings in the band structure, most prominently along H-N (indicated by red dashed circles in Fig.~\ref{fig:dft_local}(c)). By fitting the LDA+$U$ band structure to a tight-binding Hamiltonian, and restore the time-reversal symmetry, we obtained a LDA+$U$ band structure with correct time-reversal symmetry (Fig.~\ref{fig:dft_local}(d)), which is strikingly close to the open-core result in Fig.~\ref{fig:dft_local}(b). 
  
In Fig.~\ref{fig:wann_fit}, we show the comparison between the band structure from tight-binding Wannier-orbital based Hamiltonian and the original DFT results. For \pt343, we show the comparison for itinerant Ce-4$f$ calculations; while for \pd343, we show the comparison for localized Ce-4$f$ calculations (since they are most relevant). The SOCs are considered in both situations. For the former case, the fitting employs all Ce-4$f$, Pt-5$d$, and Bi-6$p$ atomic orbitals, that is,  192 Wannier orbitals in total; while for the later case, the fitting employs all Pd-4$d$, and Bi-6$p$ orbitals, that is,  108 Wannier orbitals in total. The projected Hamiltonians are symmetrized to restore all crystal symmetries as well as time-reversal symmetry. The TB Hamiltonian perfectly reproduces the DFT band structure over the whole fitting energy range, manifesting the high quality and validity of our TB Hamiltonian.

In Fig. \ref{fig:dft_dos}, we show the DFT total and projected DOS of \pd343 with and without SOC. Near Fermi level, the electronic states are dominated by Ce-5d and Bi-6p states. 

\section{Evolution of Wannier charge center}

We show our Wilson loop analysis of the band structure  of \pt343 in Fig.~\ref{fig:wanncenter}. In either calculations, we used Ce-4$f$, Pt-5$d$ and Bi-6$p$ orbitals to obtain a 192-orbital tight-binding Hamiltonian. For DFT calculations, this fitting is straightforward (using DFT eigenvalues, the same Hamiltonian produced Fig.~\ref{fig:wann_fit}). For DMFT calculations, we extract the converged self-energy at the zero-frequency $\Sigma(\omega=0)$ and add it to the DFT Hamiltonian to obtain the topological Hamiltonian)~\cite{PhysRevX.2.031008,PhysRevX.4.011006,Wang_2013}. The evolution of occupied state Wannier charge center was tracked. In either case, there are clear global gaps between the Wannier charge center lines, indicating that \pt343 is topologically trivial.

\newpage

\section{DMFT results of $\mathbf{Ce}_3\mathbf{Pd}_3\mathbf{Bi}_4$ at lower temperatures close to Fermi level}

\begin{figure*}[htp]
  \includegraphics[width=18 cm]{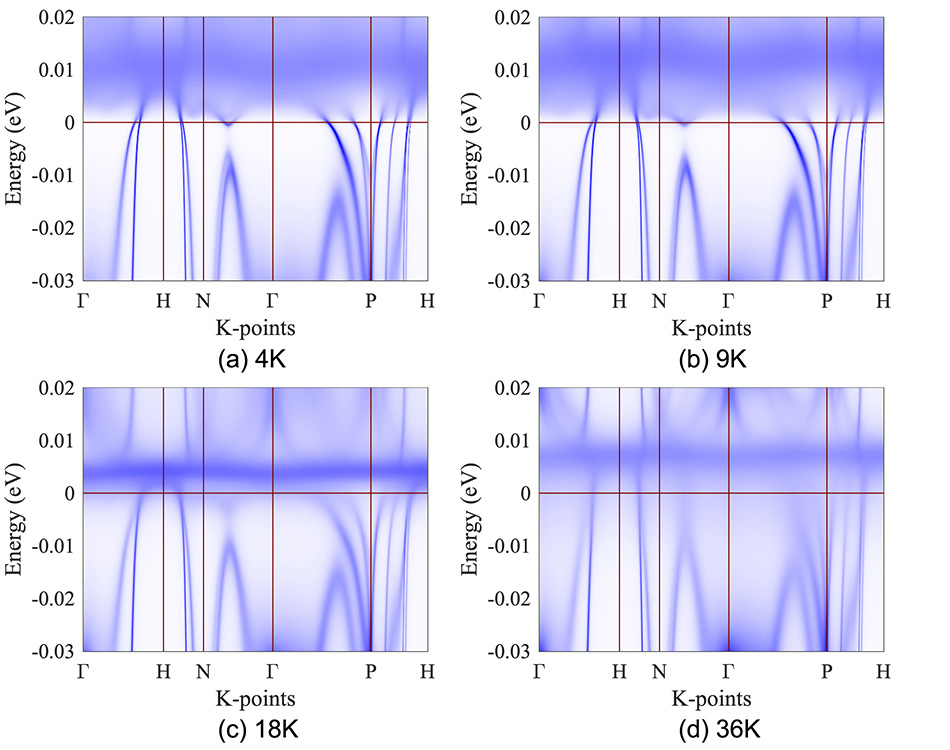}
  \caption{DFT+DMFT band structure of \pd343 close to Fermi level at temperatures below 36 K. The Fermi level is aligned at 0. \label{fig:dft_dmft}}
\end{figure*}

We show the low-temperature DMFT results of \pd343 close to the Fermi level in Fig.~\ref{fig:dft_dmft}, from which the small dip crossing the Fermi level between N and $\Gamma$ at 4 K can be clearly identified.  
\section{DFT and DMFT Fermi surfaces}

  \begin{figure*}[h]
    \subfigure[DMFT Fermi surfaces]{\includegraphics[width=16cm]{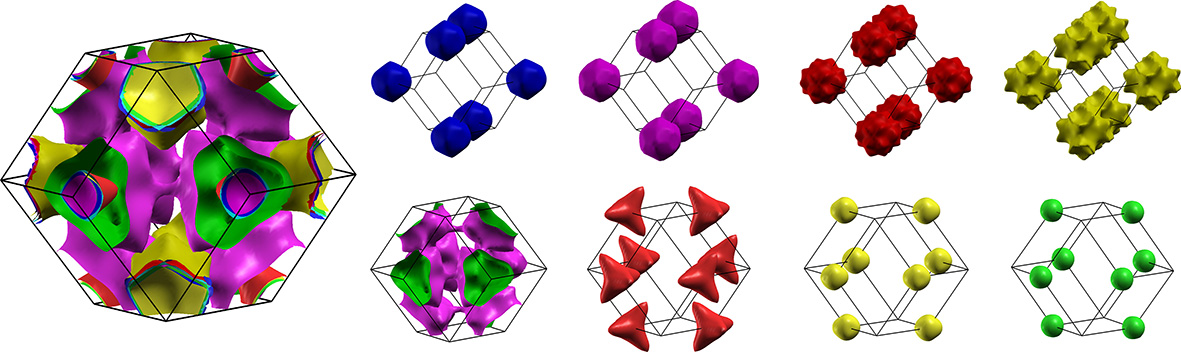}}
    \subfigure[DFT Fermi surfaces without shifting Fermi level]{\includegraphics[width=12cm]{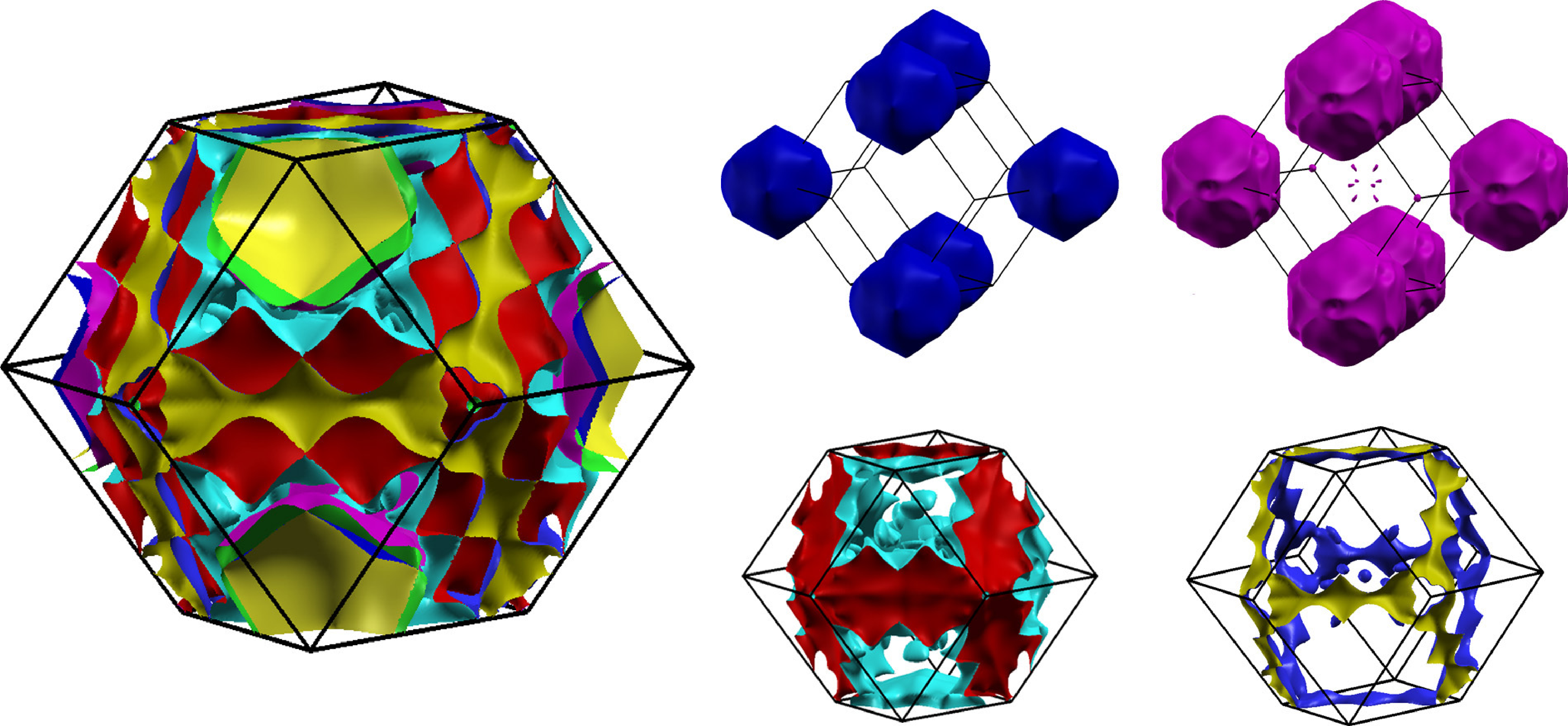}}   
    \subfigure[DFT Fermi surfaces with Fermi level shifting]{\includegraphics[width=16cm]{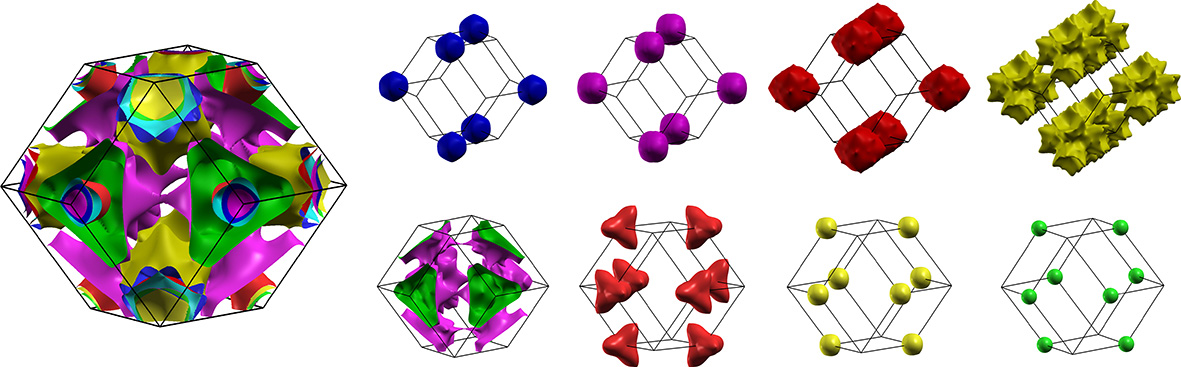}}   
    \caption{Fermi surfaces from (a) DMFT calculation, (b) DFT calculation with open-core Ce-4$f$ electrons and (c) DFT calculations with open-core Ce-4$f$ calculations but shifted by $\sim$0.2 eV (equivalent to 1 electron per f.u. doping). In (a-c) the left panels are the plot of complete Fermi surfaces, the upper right panels are the Fermi surface sheets around H point and the lower right panels are the Fermi surface sheets around P points. \label{fig:fs}}
  \end{figure*}

The Fermi surfaces of \pd343 from 4 K DMFT calculations, open-core Ce-4$f$ DFT calculations and DFT calculations with shifted Fermi level are shown in Fig.~\ref{fig:fs}. The open-core Ce-4$f$ DFT results (Fig.~\ref{fig:fs}b) are apparently very different from low temperature DMFT results (Fig.~\ref{fig:fs}(a)), but the one with shifted Fermi level (Fig.~\ref{fig:fs}(c)) seem to be topologically equivalent to the 4 K DMFT results (Fig.~\ref{fig:fs}(a)). We note  that while the DFT calculation with the shifted Fermi level can reproduce topologically equivalent Fermi surfaces, the detailed shape and size of each sheet is certainly different from DMFT results. 

\section{Details of DFT band structure of $\mathbf{Ce}_3\mathbf{Pd}_3\mathbf{Bi}_4$ close to Fermi level between $\Gamma$-H and P-H}

\begin{figure*}[h]
  \includegraphics[width=16cm]{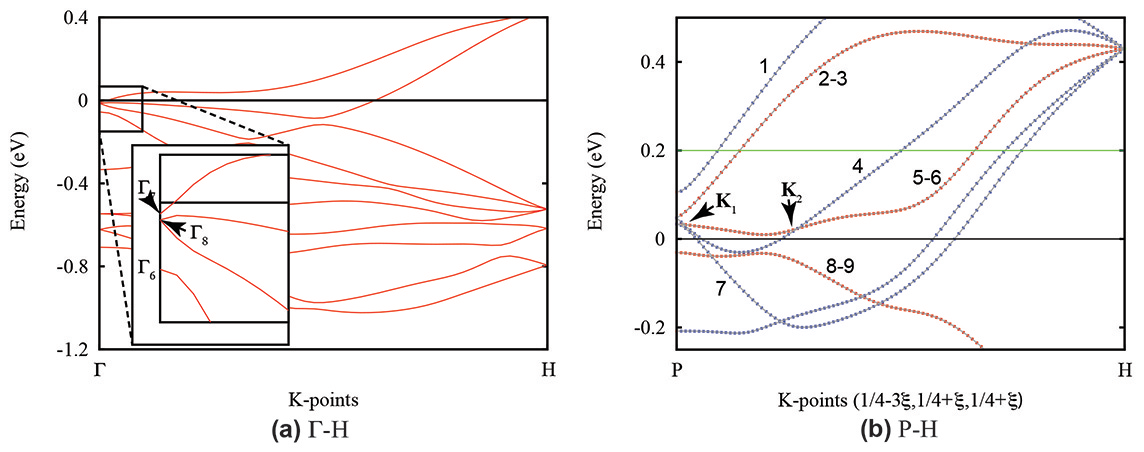}
    \caption{DFT band structure of \pd343 assuming open-core Ce-4$f$ electrons with spin-orbit coupling between (a) $\Gamma$-H and (b) P-H. The Fermi level is aligned at 0. The $\Gamma_7$, $\Gamma_8$ and $\Gamma_6$ states at $\Gamma$ are labeled in the inset of left panel. In the right panel, band 4 and doubly degenerate 5(6) crosses each other at K$_1$ and K$_2$, creating a segment of nodal line in between. \label{fig:dft_segs}}
  \end{figure*}

Some further details of \pd343 DFT band structure are shown in Fig.~\ref{fig:dft_segs}. 

\section{Comparison between $\mathbf{Ce}_3\mathbf{Pt}_3\mathbf{Bi}_4$ and $\mathbf{Ce}_3\mathbf{Pd}_3\mathbf{Bi}_4$ hybridization functions at 18K}

\begin{figure*}[h]
  \includegraphics[width=16cm]{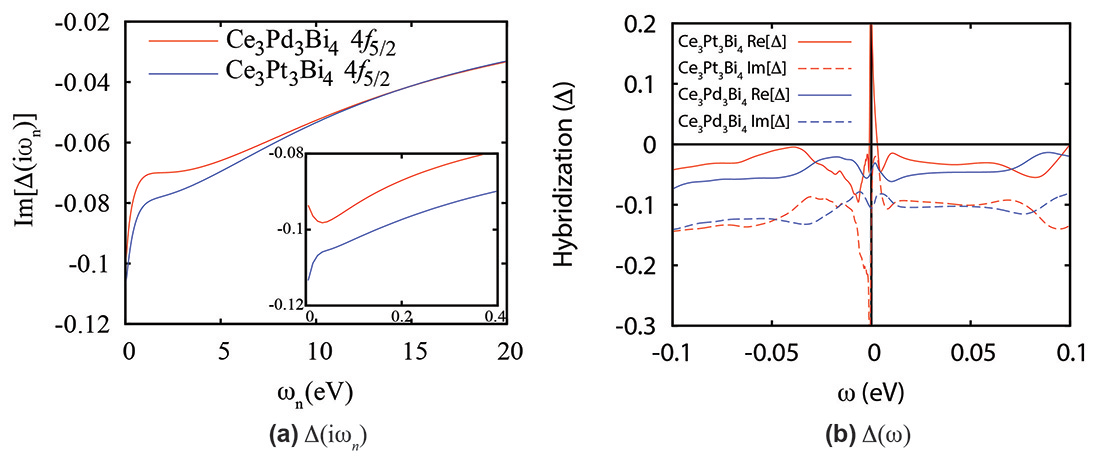}
  \caption{Comparison between hybridization functions for Ce-4$f_{5/2}$ in \pt343 and \pd343 at 18 K (a) in Matsubara frequencies and (b) in real-frequency axis. \label{fig:hyb}}
\end{figure*}

In Fig. \ref{fig:hyb}(a) we show the imaginary part of Matsubara frequency hybridization functions for Ce-4$f_{5/2}$ in \pt343 and \pd343 at 18 K. They exhibit apparent different behaviors below $\omega_n< 0.04$ eV. Correspondingly, in real-frequency axis (Fig.~\ref{fig:hyb}(b)), both the real and imaginary part of the hybridization function for \pt343 show strong peaks close to the Fermi level, which leads to the Kondo insulating behavior; while no such peaks can be found for \pd343.

\section{Convergence of Calculations at 4K}

\begin{figure*}[htp]
 \includegraphics[width=16cm]{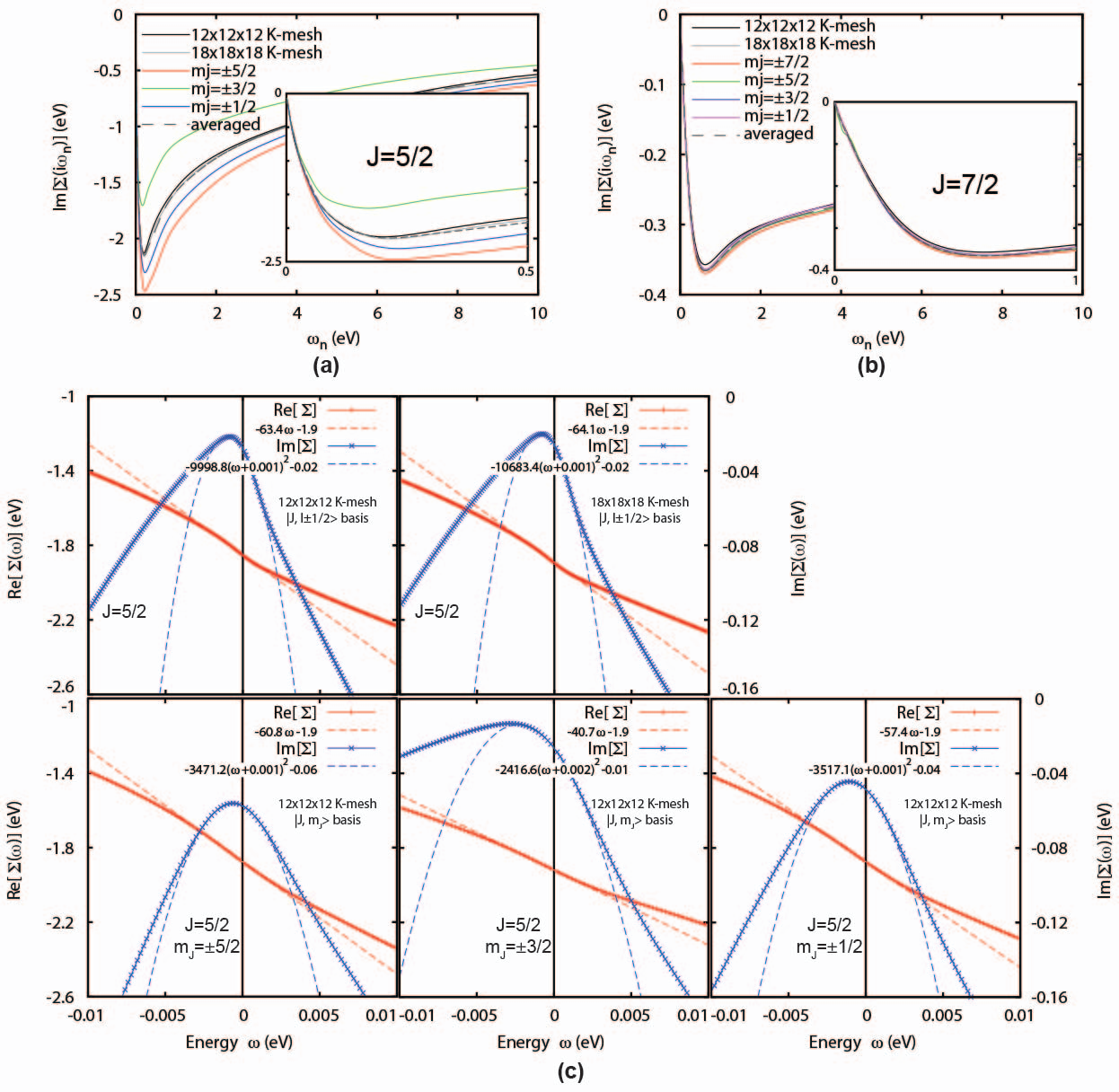}
  \caption{Self-energy $\Sigma$ obtained from different calculations at 4K. (a) Imaginary of Matsubara frequency self-energy $\mathrm{Im}[\Sigma(i\omega_n)]$ for $J$=5/2 states, (b) $\mathrm{Im}[\Sigma(i\omega_n)]$ for $J$=7/2 states, and (c) real-frequency self-energy $\Sigma(\omega)$ for $J$=5/2 states. \label{fig:sig_conv}}
\end{figure*}

\begin{figure*}[htp]
  \includegraphics[width=16cm]{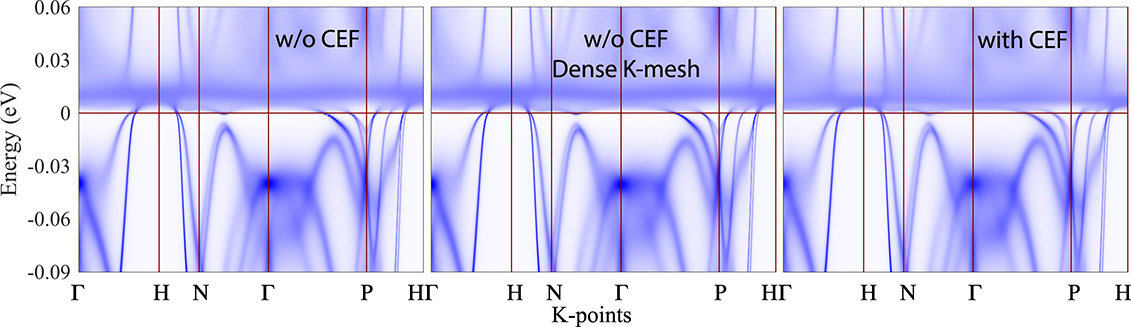}
  \caption{Comparison of momentum resolved spectrum at 4K calculated using different parameters. Left panel: the parameters employed in the main-text, i.e. $12\times12\times12$ K-mesh, $\vert j=l\pm1/2\rangle$ basis (no CEF in impurity solver), perturbation truncation order 3200. Middle panel: $18\times18\times18$ K-mesh, $\vert j=l\pm1/2\rangle$ basis, perturbation truncation order 3200. Right panel: $12\times12\times12$ K-mesh, $\vert j, m_j\rangle$ basis (explicit CEF in impurity solver), perturbation truncation order 4000. \label{fig:spec_conv}}
\end{figure*}

\begin{figure*}[htp]
  \includegraphics[width=8cm]{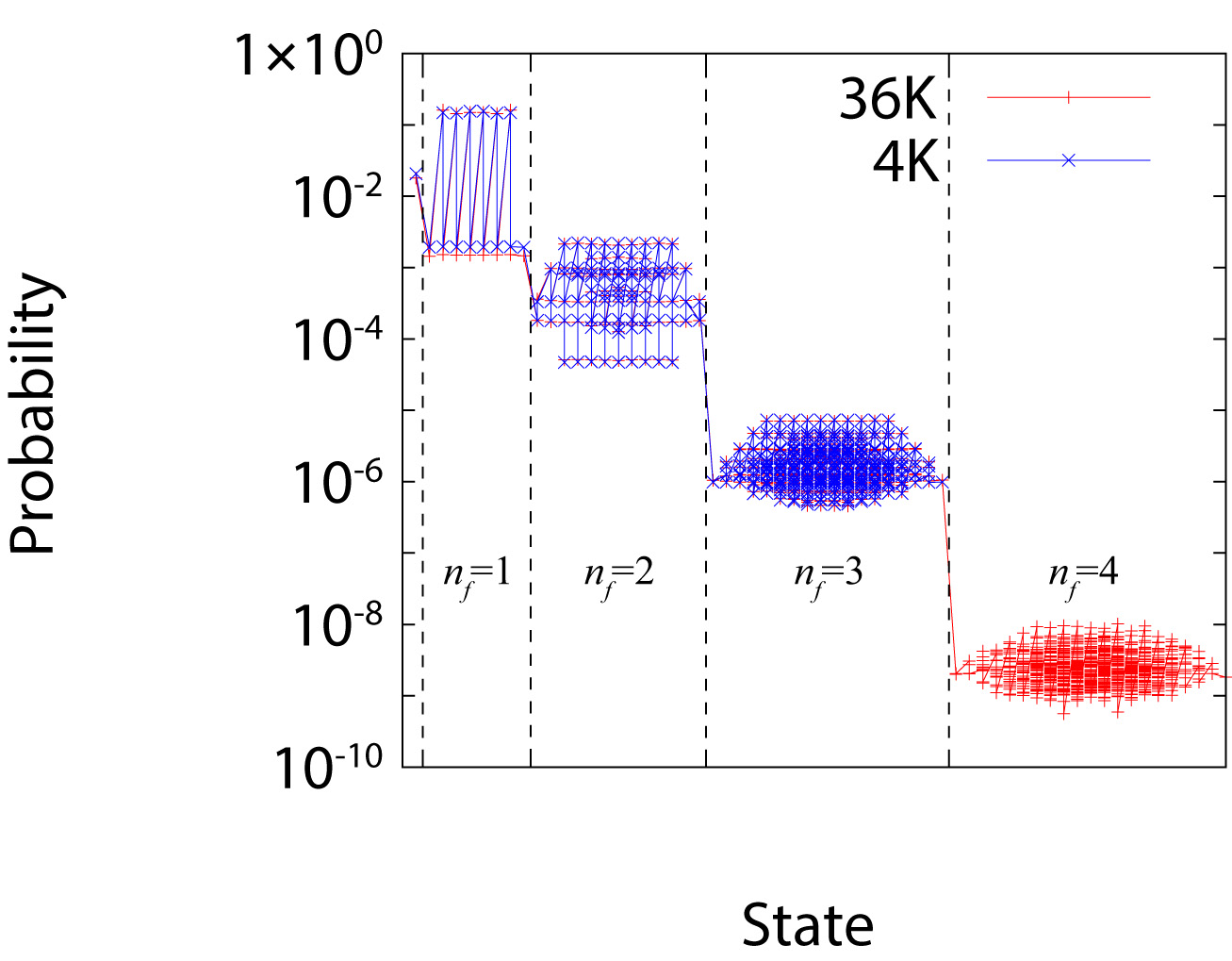}
  \caption{Probability of different $n_f$ occupations.
  \label{fig:prob}}
\end{figure*}

In order to show that our calculations at 4K is already converged with respect to the K-mesh size, we have also performed calculations with denser $18\times18\times18$ K-mesh, starting from the converged result with $12\times12\times12$ K-mesh. The calculation was done with 25 DMFT iterations (each DMFT iteration contains 1 CT-QMC iteration with $2\times10^9$ QMC steps on 144 cores), and the resulting self-energy was also taken from the average of the last 5-iterations. We plot the imaginary part of Matsubara frequency self-energy $\mathrm{Im}[\Sigma(i\omega_n)]$ in Fig. \ref{fig:sig_conv}(a-b), and the real-frequency self-energy $\Sigma(\omega)$ in Fig. \ref{fig:sig_conv}c. It is apparent that the difference between the resulting Matsubara-frequency self-energies $\mathrm{Im}[\Sigma(i\omega_n)]$ [grey solid lines in Fig. \ref{fig:sig_conv}(a-b)] and $12\times12\times12$ K-mesh results [black solid lines in Fig. \ref{fig:sig_conv}(a-b)] is negligible. In addition, the fitting of the real-frequency self-energy yields mass-enhancement of $\sim 65.1$, whereas the $12\times12\times12$ K-mesh result is $\sim 64.7$. Therefore, these results are almost identical, and $12\times12\times12$ K-mesh should be sufficient to converge the calculations.

\begin{table}
 \begin{tabular}{c|c|c|c|c|c|c|c|c|c|c|c|c|c}
 \hline\hline
 \multicolumn{14}{c}{$\omega$=0} \\
 \hline
0.473 & 0.006 & 0.015 & 0.007 & 0.004 & 0.002 & 0.007 & 0.006 & 0.003 & 0.008 & 0.004 & 0.009 & 0.002 & 0.016\\
0.006 & 0.483 & 0.004 & 0.006 & 0.004 & 0.004 & 0.006 & 0.001 & 0.007 & 0.006 & 0.004 & 0.007 & 0.005 & 0.006\\
0.015 & 0.004 & 0.455 & 0.006 & 0.006 & 0.007 & 0.002 & 0.002 & 0.002 & 0.019 & 0.003 & 0.007 & 0.004 & 0.008\\
0.007 & 0.006 & 0.006 & 0.455 & 0.004 & 0.014 & 0.008 & 0.004 & 0.007 & 0.002 & 0.019 & 0.003 & 0.003 & 0.001\\
0.004 & 0.004 & 0.006 & 0.004 & 0.483 & 0.005 & 0.005 & 0.005 & 0.007 & 0.004 & 0.006 & 0.007 & 0.000 & 0.007\\
0.002 & 0.004 & 0.007 & 0.014 & 0.005 & 0.473 & 0.016 & 0.002 & 0.009 & 0.004 & 0.008 & 0.003 & 0.006 & 0.007\\
0.007 & 0.006 & 0.002 & 0.008 & 0.005 & 0.016 & 0.830 & 0.003 & 0.006 & 0.003 & 0.008 & 0.005 & 0.020 & 0.004\\
0.006 & 0.001 & 0.002 & 0.004 & 0.005 & 0.002 & 0.003 & 0.868 & 0.004 & 0.014 & 0.005 & 0.010 & 0.004 & 0.020\\
0.003 & 0.007 & 0.002 & 0.007 & 0.007 & 0.009 & 0.006 & 0.004 & 0.821 & 0.004 & 0.010 & 0.001 & 0.010 & 0.005\\
0.008 & 0.006 & 0.019 & 0.002 & 0.004 & 0.004 & 0.003 & 0.014 & 0.004 & 0.838 & 0.003 & 0.010 & 0.005 & 0.008\\
0.004 & 0.004 & 0.003 & 0.019 & 0.006 & 0.008 & 0.008 & 0.005 & 0.010 & 0.003 & 0.839 & 0.004 & 0.013 & 0.003\\
0.009 & 0.007 & 0.007 & 0.003 & 0.007 & 0.003 & 0.005 & 0.010 & 0.001 & 0.010 & 0.004 & 0.821 & 0.004 & 0.006\\
0.002 & 0.005 & 0.004 & 0.003 & 0.000 & 0.006 & 0.020 & 0.004 & 0.010 & 0.005 & 0.013 & 0.004 & 0.868 & 0.003\\
0.016 & 0.006 & 0.008 & 0.001 & 0.007 & 0.007 & 0.004 & 0.020 & 0.005 & 0.008 & 0.003 & 0.006 & 0.003 & 0.830\\
\hline
 \multicolumn{14}{c}{$\omega\rightarrow\infty$}\\
 \hline
0.542 & 0.001 & 0.004 & 0.001 & 0.003 & 0.000 & 0.001 & 0.002 & 0.002 & 0.003 & 0.001 & 0.010 & 0.003 & 0.006\\
0.001 & 0.535 & 0.001 & 0.003 & 0.001 & 0.003 & 0.003 & 0.002 & 0.000 & 0.002 & 0.001 & 0.003 & 0.009 & 0.002\\
0.004 & 0.001 & 0.537 & 0.000 & 0.003 & 0.001 & 0.001 & 0.001 & 0.001 & 0.001 & 0.001 & 0.003 & 0.003 & 0.008\\
0.001 & 0.003 & 0.000 & 0.537 & 0.001 & 0.004 & 0.008 & 0.003 & 0.003 & 0.001 & 0.001 & 0.001 & 0.001 & 0.001\\
0.003 & 0.001 & 0.003 & 0.001 & 0.535 & 0.001 & 0.002 & 0.009 & 0.003 & 0.001 & 0.002 & 0.000 & 0.002 & 0.003\\
0.000 & 0.003 & 0.001 & 0.004 & 0.001 & 0.542 & 0.006 & 0.003 & 0.010 & 0.001 & 0.003 & 0.002 & 0.002 & 0.001\\
0.001 & 0.003 & 0.001 & 0.008 & 0.002 & 0.006 & 0.873 & 0.001 & 0.005 & 0.001 & 0.006 & 0.001 & 0.003 & 0.000\\
0.002 & 0.002 & 0.001 & 0.003 & 0.009 & 0.003 & 0.001 & 0.870 & 0.001 & 0.002 & 0.001 & 0.001 & 0.001 & 0.003\\
0.002 & 0.000 & 0.001 & 0.003 & 0.003 & 0.010 & 0.005 & 0.001 & 0.868 & 0.000 & 0.003 & 0.001 & 0.001 & 0.001\\
0.003 & 0.002 & 0.001 & 0.001 & 0.001 & 0.001 & 0.001 & 0.002 & 0.000 & 0.868 & 0.000 & 0.003 & 0.001 & 0.006\\
0.001 & 0.001 & 0.001 & 0.001 & 0.002 & 0.003 & 0.006 & 0.001 & 0.003 & 0.000 & 0.868 & 0.000 & 0.002 & 0.001\\
0.010 & 0.003 & 0.003 & 0.001 & 0.000 & 0.002 & 0.001 & 0.001 & 0.001 & 0.003 & 0.000 & 0.868 & 0.001 & 0.005\\
0.003 & 0.009 & 0.003 & 0.001 & 0.002 & 0.002 & 0.003 & 0.001 & 0.001 & 0.001 & 0.002 & 0.001 & 0.870 & 0.001\\
0.006 & 0.002 & 0.008 & 0.001 & 0.003 & 0.001 & 0.000 & 0.003 & 0.001 & 0.006 & 0.001 & 0.005 & 0.001 & 0.873\\
\hline
 \end{tabular}
 \caption{Absolute value of impurity level matrix.}
 \label{tab:imp}
\end{table}

In addition, we have also performed calculations with $\vert j, m_j\rangle$ basis to identify the effect of CEF. When performing such calculations, we have increased the perturbation order to 4000. These results are also illustrated in Fig. \ref{fig:sig_conv}. For $j=7/2$ states, we notice that the Matsubara-frequency self-energy $\mathrm{Im}[\Sigma(i\omega_n)]$ for all $\vert 7/2, m_j\rangle$ ($m_j=\pm1/2, \pm3/2, \pm5/2, \pm7/2$) states are similar to one another, and the differences are negligible. For $j=5/2$ states, $\mathrm{Im}[\Sigma(i\omega_n)]$ for $m_j=\pm1/2, \pm3/2, \pm5/2$ states have similar shapes. Moreover, the real-frequency self-energy yields similar mass-enhancement ($\sim$ 62, 42, 58 for $m_j=\pm5/2, \pm3/2, \pm1/2$ states, respectively). The converged impurity levels at 4K are 0.542 eV, 0.535 eV and 0.537 eV for $j=5/2$ states; as well as 0.873 eV, 0.870 eV, 0.868 eV, and 0.868 eV for $j=7/2$ states. Thus the renormalized CEF is negligible, and therefore we could ignore it at the impurity-solver level. As a result, the $\mathbf{k}$-resolved spectrum of these calculations are similar, and no apparent difference can be observed (Fig. \ref{fig:spec_conv}).

We also show the QMC weight of the many-body states for different occupations $n_f$ (Fig.\ref{fig:prob}). For 36K calculation, we show the quantity up to $n_f=4$, whereas for 4K calculations, we calculate up to $n_f=3$. The order of magnitude of these weights do not change from 36K to 4K, and weights of $n_f=4$ states are at least 2 orders of magnitude smaller than those of $n_f=3$ states, and approximately 6 orders of magnitude smaller than those of $n_f=1$ states. At 4K, the accumulated weight of $n_f=0$, 1, 2, and 3 states are 0.0199, 0.916, 0.063, and 0.0007, respectively. These sum up to $\sim$ 0.9996, therefore it is safe for us to restrict the occupation within [0, 3] in the impurity solver. 

Finally, we examine the off-diagonal elements of the full impurity level matrix. We list the absolute value of the full matrix in TAB. \ref{tab:imp}. We observe that most of the off-diagonal matrix elements are order of several meVs, whereas the largest ones are $\sim$ 20 meV at zero-frequency. When $\omega\rightarrow\infty$, these off-diagonal matrix elements are even smaller, with maximum around $\sim$ 10 meV. It is therefore safe to ignore the off-diagonal matrix elements in the impurity solver. Furthermore, since the off-diagonal matrix elements are negligible, we ignore the off-diagonal hybridization terms. Therefore, we do not observe sign problem in our calculations~\cite{method:dmft2,RevModPhys.83.349}, and the averaged signs at all temperatures are +1.

\bibliography{cbp343}